\documentclass[twocolumn]{aastex62}

\graphicspath{{Figures/}}

\received{}
\revised{}
\accepted{}
\submitjournal{ApJ}

\shorttitle{Vortex Fiber Nulling}
\shortauthors{Ruane et al.}

\begin{document}

\title{\large Efficient spectroscopy of exoplanets at small angular separations with vortex fiber nulling}

\correspondingauthor{Garreth Ruane}
\email{gruane@caltech.edu}

\author[0000-0003-4769-1665]{Garreth Ruane}
\altaffiliation{NSF Astronomy and Astrophysics Postdoctoral Fellow}
\affil{Department of Astronomy, California Institute of Technology, 1200 E. California Blvd., Pasadena, CA 91125, USA}

\author[0000-0002-4361-8885]{Ji Wang}
\affil{Department of Astronomy, California Institute of Technology, 1200 E. California Blvd., Pasadena, CA 91125, USA}

\author[0000-0002-8895-4735]{Dimitri Mawet}
\affil{Department of Astronomy, California Institute of Technology, 1200 E. California Blvd., Pasadena, CA 91125, USA}
\affil{Jet Propulsion Laboratory, California Institute of Technology, 4800 Oak Grove Dr., Pasadena, CA 91109, USA}

\author{Nemanja~Jovanovic}
\affil{Department of Astronomy, California Institute of Technology, 1200 E. California Blvd., Pasadena, CA 91125, USA}

\author{Jacques-Robert~Delorme}
\affil{Department of Astronomy, California Institute of Technology, 1200 E. California Blvd., Pasadena, CA 91125, USA}

\author{Bertrand~Mennesson}
\affil{Jet Propulsion Laboratory, California Institute of Technology, 4800 Oak Grove Dr., Pasadena, CA 91109, USA}

\author[0000-0001-5299-6899]{J.~Kent Wallace}
\affil{Jet Propulsion Laboratory, California Institute of Technology, 4800 Oak Grove Dr., Pasadena, CA 91109, USA}

\begin{abstract}
Instrumentation designed to characterize potentially habitable planets may combine adaptive optics and high-resolution spectroscopy techniques to achieve the highest possible sensitivity to spectral signs of life. Detecting the weak signal from a planet containing biomarkers will require exquisite control of the optical wavefront to maximize the planet signal and significantly reduce unwanted starlight. We present an optical technique, known as vortex fiber nulling (VFN), that allows polychromatic light from faint planets at extremely small separations from their host stars ($\lesssim\lambda/D$) to be efficiently routed to a diffraction-limited spectrograph via a single-mode optical fiber, while light from the star is prevented from entering the spectrograph. VFN takes advantage of the spatial selectivity of a single-mode fiber to isolate the light from close-in companions in a small field of view around the star. We provide theoretical performance predictions of a conceptual design and show that VFN may be utilized to characterize planets detected by radial velocity (RV) instruments in the infrared without knowledge of the azimuthal orientation of their orbits. Using a spectral template-matching technique, we calculate an integration time of $\sim$400, $\sim$100, and $\sim$30 hr for Ross 128 b with Keck, the Thirty Meter Telescope (TMT), and the Large Ultraviolet/Optical/Infrared (LUVOIR) Surveyor, respectively. 
\end{abstract}


\section{Introduction} \label{sec:intro}

Perhaps the only practical pathway for detecting biosignatures with ground-based telescopes is to obtain high-resolution spectra of planets orbiting the nearest M-dwarf stars \citep{Riaud2007,Snellen2015,Wang2017_HDCI}. The recent discoveries of Proxima~Centauri~b \citep{proxcenb} and Ross~128~b \citep{Bonfils2017} are examples of what may be a plentiful sample of tantalizing targets. However, these planets are out of reach for current high-contrast imaging instruments because of the extremely small angular separation ($<$37 and $<$15~mas, respectively) and flux ratio ($\sim10^{-8}$-$10^{-7}$) between the planets and their host stars. The necessary inner working angle in both cases is smaller than the angular resolution of a 10~meter telescope in the infrared. Moreover, the wavefront control precision needed to sufficiently suppress unwanted starlight is $\sim$100$\times$ better than provided by current state-of-the-art adaptive optics (AO) systems. The discovery of life on these worlds via imaging spectroscopy may therefore need to wait for next-generation extreme AO on a giant segmented mirror ground-based telescopes, such as the Planetary Systems Imager (PSI) on the Thirty Meter Telescope (TMT), or large-aperture space telescopes such as the Large Ultraviolet/Optical/Infrared Surveyor \citep[LUVOIR;][]{Bolcar2016,Pueyo2017}. Even with a primary mirror diameter of TMT ($D$=30~m), this application requires an instrument that provides robust rejection of light from the star whose photon noise contribution overwhelms the relatively few photons from a planet with an angular separation of $\lesssim\lambda/D$ \citep{Kawahara2014}. 

\begin{figure*}[t!]
    \centering
    \includegraphics[width=0.8\linewidth]{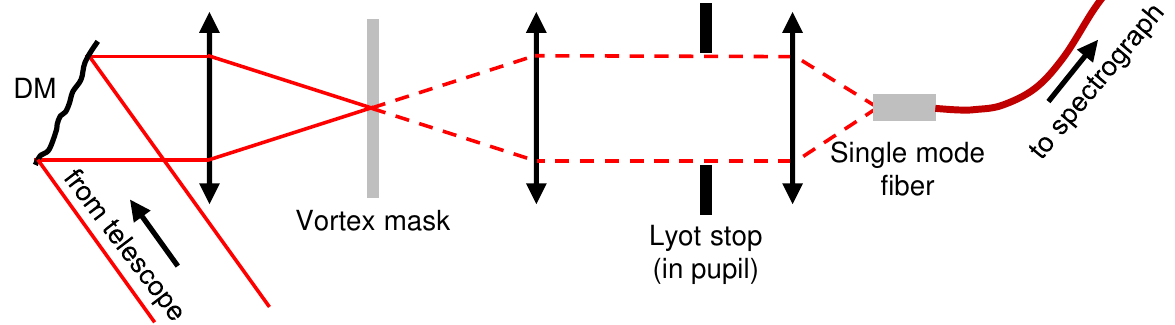}
    \caption{Schematic of the optical system. The wavefront from the telescope is flattened by a deformable mirror (DM) and the beam is focused such that the star is aligned at the center of the vortex phase mask in the focal plane. Some of the starlight is blocked at the Lyot stop in the downstream pupil. The single mode fiber (SMF) is centered at the geometric image of the star and is 2.5$\times$ larger than $\lambda F^\#$ at the fiber to simultaneously capture off-axis planet light. Since the vortex mask imparts an azimuthal phase ramp given by $\exp(\pm i\theta)$, the reimaged stellar field is orthogonal to the fundamental mode of the SMF and is therefore rejected.}
    \label{fig:OptDiagram}
\end{figure*}

Here, we present the vortex fiber-nulling (VFN) concept, which imparts an optical vortex phase pattern on the focused starlight \citep{Foo2005,Mawet2005} causing it to be rejected by a single-mode fiber (SMF) in the subsequent focal plane in a process akin to fiber-nulling interferometry \citep{Haguenauer2006,Mennesson2006,Por2018}. The SMF feeds light from planets at angular separations $\sim\lambda/D$ into a high-resolution spectrograph (R~$=\lambda/\Delta\lambda\approx$~100,000). The spectrum is then analyzed using high-dispersion coronagraphy methods \citep{SparksFord2002,Riaud2007,Snellen2015,Wang2017_HDCI,Mawet2017_HDCII,Lovis2017} to reveal close-in planetary mass companions in small field of view around the star and identify molecular species in their atmospheres. 

Our optical design may potentially be used to detect and characterize planets as close as $\sim$0.4~$\lambda/D$ from the star over a wide spectral range and does not require precise prior knowledge of the orientation of the planet's orbit. The VFN method will unlock the potential to discover and characterize low-mass exoplanets in the near future at the W.M. Keck Observatory with minor modifications to the upcoming Keck Planet Imager and Characterizer (KPIC) instrument \citep{Mawet2016_KPIC,Mawet2017_KPIC}. Using Ross~128~b as an example, we compute the integration time needed to detect potential signs of life in the atmospheres of terrestrial planets orbiting in the habitable zone of nearby M stars with Keck, TMT, and LUVOIR. The feasibility of such observations is dependent on the AO system's ability to control a select few low-order wavefront error modes, namely tip-tilt and coma, and is relatively insensitive to mid and high spatial frequency aberrations.\\ 
\\

\section{VFN Concept}


Vortex fiber nullers are designed to be sensitive to planet light at small angular separations while suppressing unwanted starlight. Figure \ref{fig:OptDiagram} shows a schematic of a VFN instrument consisting of a deformable mirror (DM) for wavefront control, followed by a vortex phase mask in the focal plane with complex transmittance $\exp(\pm i\theta)$, where $\theta$ is the polar angle in the focal plane. The starlight is centered on the phase singularity of the focal plane mask. A VFN instrument is different than a vortex coronagraph in that the starlight is then re-imaged onto a co-aligned SMF, which rejects it.
That is, the coupling efficiency of the field, $f(r,\theta)$, at the fiber tip is nulled if the overlap integral with the fiber mode is zero; i.e.,
\begin{equation}
    \int \psi(r) f(r,\theta) dA = 0,
    \label{eqn:overlap}
\end{equation}
where $\psi(r)$ is the fundamental mode of the SMF. For common SMFs, the mode may be approximated by $\psi(r)=\exp[-(2r/D_f)^2]$, where $D_f$ is the mode field diameter. The starlight has an azimuthally varying phase term of the form $\exp(\pm i\theta)$, which is orthogonal to the uniform phase of $\psi(r)$ and therefore prevents the starlight from coupling into the SMF. In fact, Eqn.~\ref{eqn:overlap} is zero if the stellar field is of the form $f(r,\theta)=f_r(r)\exp(il\theta)$ with a nonzero integer value of~$l$. Furthermore, we confirmed numerically that an azimuthally dependent fiber mode, $\psi(r,\theta)$, may also reject the $\exp(il\theta)$ term for a variety of fold symmetries, including the hexagonal mode of a photonic crystal fiber \citep{Birks1997}. 

\begin{figure}[t]
    \centering
    \includegraphics[height=0.8\linewidth,trim={0 -1.5mm 0 0},clip]{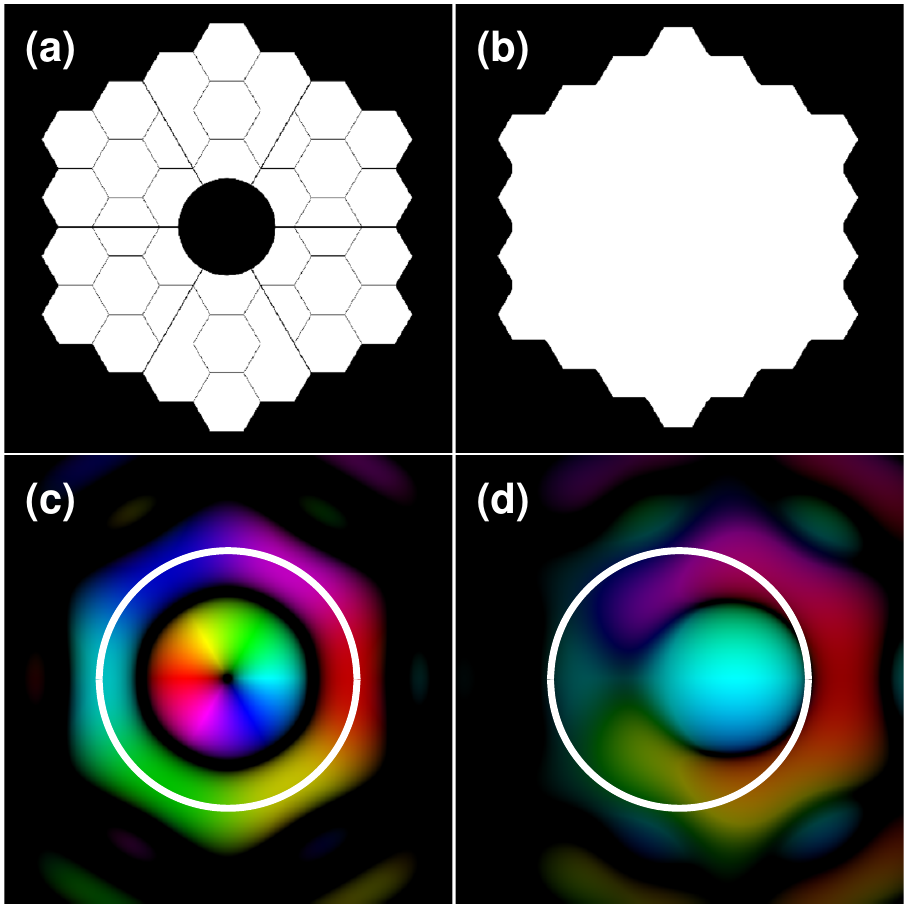}
    \includegraphics[height=0.8\linewidth]{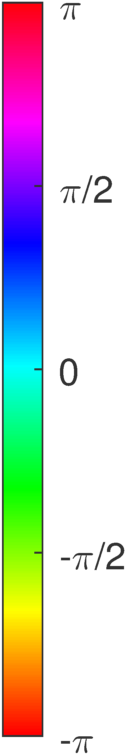}
    \caption{(a)~The Keck pupil. (b)~The Lyot stop. (c)~The amplitude (value) and phase (hue) of the stellar field in the central 4$\times$4~$\lambda F^{\#}$ region. (d)~Same as~(c), but for a planet with an angular separation of $\lambda/D$. The white circle indicates the mode field diameter of the single-mode fiber ($D_f=2.5~\lambda F^{\#}$) that will cancel the starlight and couple a fraction of the off-axis planet light.}
    \label{fig:modes}
\end{figure}

Assuming that there is no wavefront error in the system, light from an on-axis point source does not enter the SMF regardless of the $F^\#$ of the focusing optic. We therefore align the SMF to the position of the star and choose the mode field diameter and focusing optic such that $D_f=2.5\lambda F^\#$ 
to maximize planet coupling efficiency. In this configuration, the starlight is rejected by the SMF, and the light from planets at small angular separations ($\lesssim\lambda/D$) is partially coupled.

Figure \ref{fig:modes} illustrates the SMF spatial filtering mechanism with Keck's segmented, noncircular, obstructed aperture (Fig.~\ref{fig:modes}a). The Lyot stop (Fig.~\ref{fig:modes}b) is designed to only block starlight that is diffracted outside of the geometric pupil in order to reduce the amount of starlight in the image plane without paying a significant throughput penalty. The azimuthal phase dependence of the stellar field at the fiber tip (Fig.~\ref{fig:modes}c) causes Eqn. \ref{eqn:overlap} to compute to zero and therefore no starlight can propagate into the SMF. On the other hand, $\sim$20\% of the light from a planet at an angular separation of $\sim$0.9$\lambda/D$ (Fig.~\ref{fig:modes}d) makes its way into the fiber and transmits to the spectrograph. 

Whereas traditional fiber-nulling interferometers create destructive interference between sub-apertures \citep{Haguenauer2006}, VFN has the advantage of using the entirety of the telescope aperture and does not require baseline modulation \citep{Mennesson2006} to detect sources at all possible azimuthal orientations. The VFN starlight cancellation mechanism also does not depend on the wavelength nor the shape of the pupil. The coupling efficiency of the stellar beam is zero regardless of the shape of the Lyot stop; in fact, the Lyot stop may be removed from the system. We have opted to include a Lyot stop because it offers the practical benefit of preventing 70\% of the starlight from reaching the image plane, assuming a flat wavefront, while only reducing the planet throughput by 5\%.

\begin{figure}[t]
    \centering
    \includegraphics[width=0.9\linewidth]{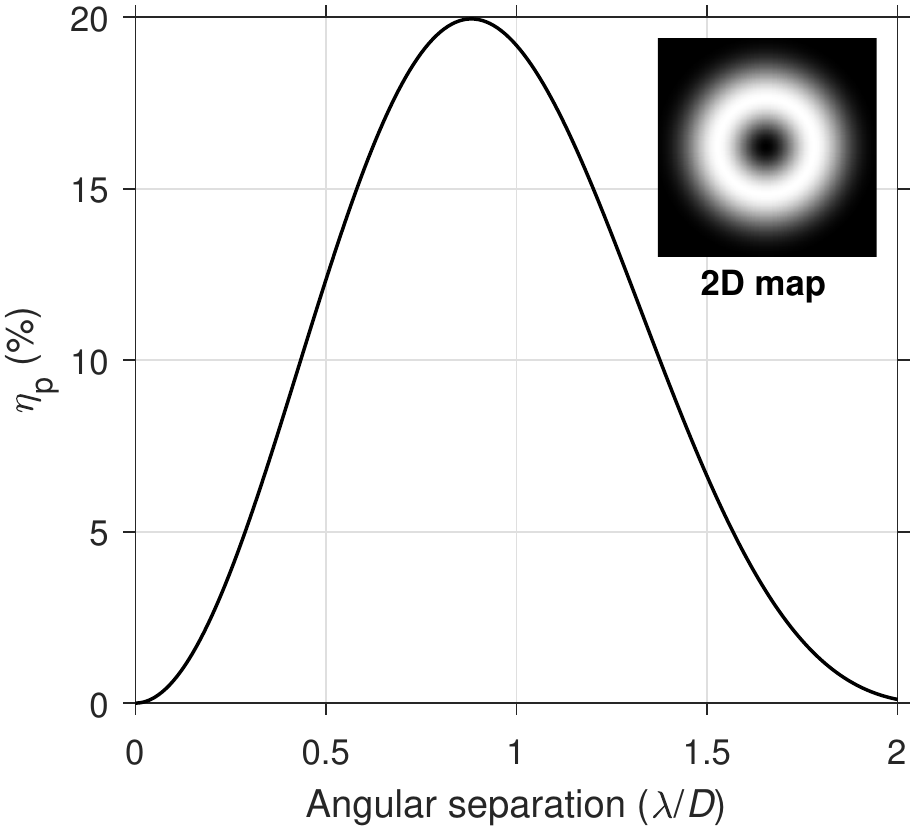}
    \caption{Theoretical planet throughput into an SMF, located on the optical axis with a mode field diameter of $D_f=2.5~\lambda F^\#$, downstream of a vortex phase mask with complex transmittance $\exp(\pm i\theta)$. This calculation accounts for diffractive losses owing to the phase mask and Lyot stop as well as the SMF coupling efficiency. With the star centered on the optical axis, light from point sources in an annular region around the star will transmit into the spectrograph. The inset shows the 2D throughput map for planets in a 4$\times$4~$\lambda/D$ field of view. The throughput only changes by 1-2\% for a range of possible telescope apertures.}
    \label{fig:thpt}
\end{figure}

The fraction of the planet light that couples into the SMF, $\eta_p$, is shown in Fig. \ref{fig:thpt}. Using the on-axis fiber mitigates the need to know the planet's precise position in advance, since the throughput does not depend on azimuth. With the star and SMF co-aligned, light from point sources in an annular region around the star will couple into the SMF (see Fig. \ref{fig:thpt}, inset). However, the mode entering the spectrograph will be $\psi(r)$ regardless of the planet location, enabling temporally stable and diffraction-limited spectroscopy.

\begin{figure*}[t!]
    \centering
    \includegraphics[width=0.325\linewidth]{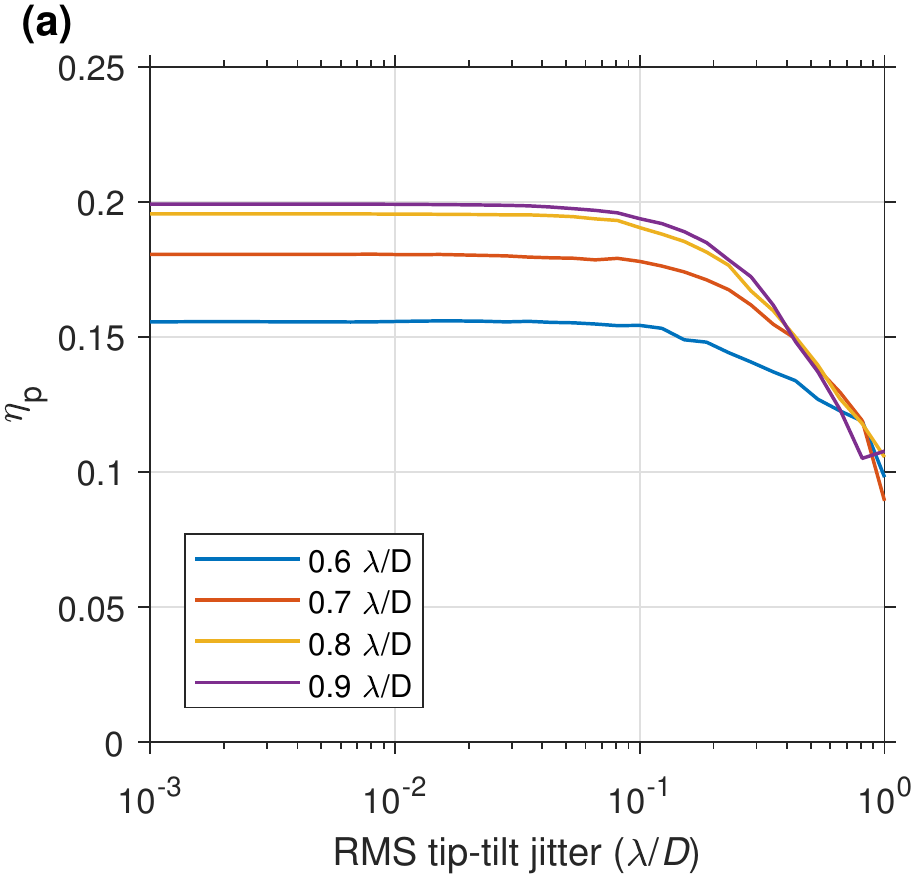}
    \includegraphics[width=0.325\linewidth]{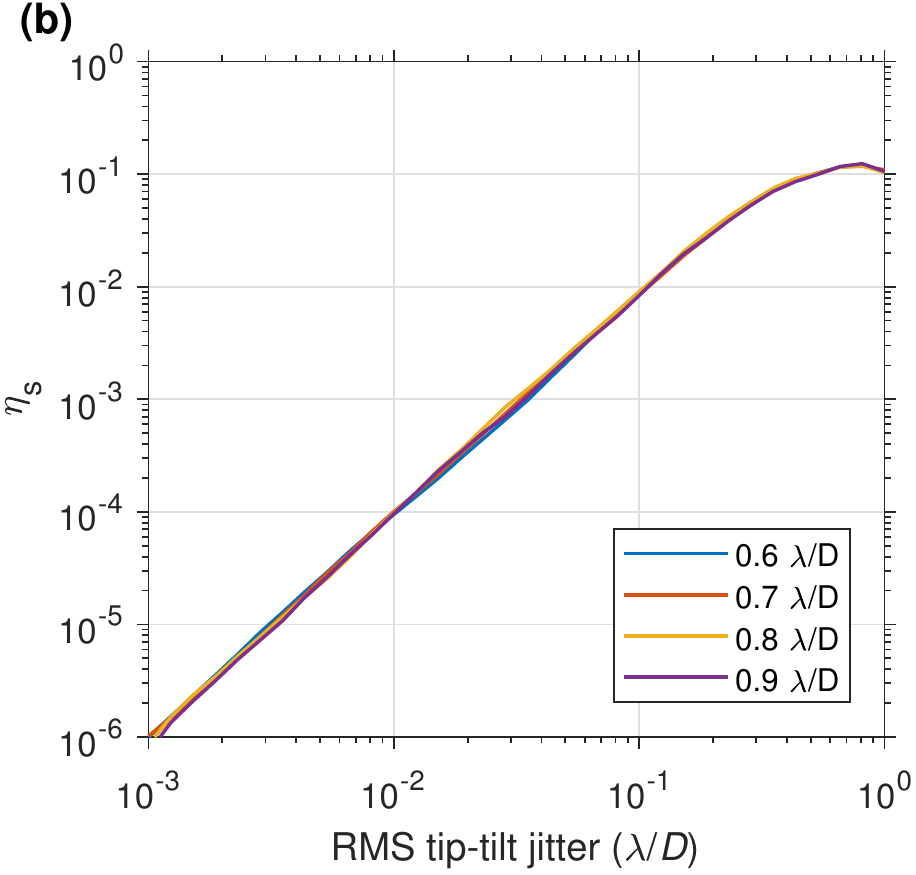}
    \includegraphics[width=0.325\linewidth]{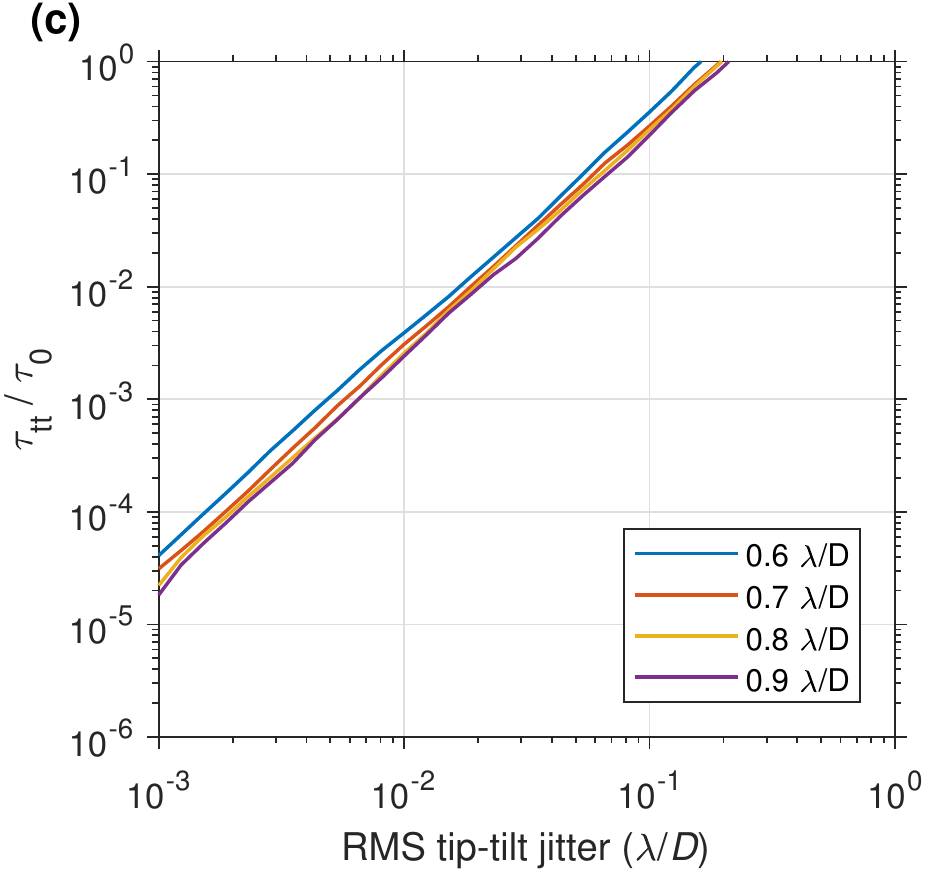}    
    \caption{Fraction of (a) planet and (b) star light transmitted through the vortex mask, Lyot stop, and SMF at the central wavelength as a function of rms tip-tilt jitter for planets at angular separations of 0.6-0.9~$\lambda/D$. (c) The relative integration time in the photon-noise-limited regime. VFN significantly reduces integration time provided tip-tilt errors are sufficiently controlled. These numerical calculations assume tip-tilt errors are normally distributed and that the star and planet are point sources.}
    \label{fig:jitter}
\end{figure*}

\section{Theoretical performance}

In this section, we predict the performance of VFN in realistic noise environments. We write the signal from the planet and star that enters the spectrograph as $S_p = \eta_p \Phi_p \tau \Delta\lambda A q T$ and $S_s = \eta_s \Phi_s \tau \Delta\lambda A q T$, where $\eta_p$ and $\eta_s$ are the planet and star throughputs of the VFN, $\Phi_p$ and $\Phi_s$ are the flux owing to the planet and star (photons per unit area per unit time per unit wavelength at the primary mirror), $\tau$ is the integration time, $\Delta\lambda$ is the full spectral bandwidth, $A$ is the collecting area of the telescope, $q$ is the quantum efficiency of the detector, and $T$ is the transmission of the instrument describing losses that affect the star and planet equally. 

When limited by stellar photon noise, the signal-to-noise ratio (SNR) per spectral channel is
\begin{equation}
\mathrm{SNR} = \frac{S_p}{\sqrt{S_s}}=\frac{\eta_p }{\sqrt{\eta_s}}\frac{\Phi_p}{\sqrt{ \Phi_s}}\sqrt{\frac{\tau \lambda_0 A q T}{R}},
\label{eq:snr1}
\end{equation}
where $\lambda$ is the wavelength and $R$ is the spectral resolution. 
Solving for the integration time to achieve a given SNR:
\begin{equation}
\tau = \frac{\eta_s}{\eta_p^2} \frac{R}{\lambda}\frac{(\mathrm{SNR})^2}{\epsilon^2\Phi_s  A q T}=\frac{\eta_s}{\eta_p^2} \tau_0,
\end{equation}
where $\epsilon = \Phi_p/\Phi_s$ is the flux ratio between the planet and the star and
\begin{equation}
\tau_0 = \frac{R}{\lambda}\frac{(\mathrm{SNR})^2}{\epsilon^2\Phi_s  A q T}.
\end{equation}
The objective of VFN is to minimize the integration time to detect molecules in a planet's atmosphere. In the remainder of this section, we confirm that the VFN approach offers significant reductions in integration time when tip-tilt and coma wavefront errors are sufficiently controlled. We also discuss the effects of partially resolved stars, background noise, and detector noise.

\subsection{Tip-tilt jitter}

On ground-based telescopes, the performance of the proposed system will likely be limited by the ability of current AO systems to accurately sense and correct tip-tilt errors at kHz rates. We assume in the following that atmospheric dispersion is compensated to high precision \citep[e.g. using methods demonstrated in][]{Pathak2018} and is therefore negligible. Figure~\ref{fig:jitter} shows $\eta_p$ and $\eta_s$ as a function of tip-tilt jitter for various angular separations of the planet. The planet throughput (Fig. \ref{fig:jitter}a) is relatively insensitive to tip-tilt jitter $<$0.1~$\lambda/D$ rms. 
On the other hand, the fraction of starlight that leaks into the SMF in the VFN configuration (Fig. \ref{fig:jitter}b) may be approximated as
\begin{equation}
    \eta_s=\left(\sigma\frac{D}{\lambda}\right)^2,
\end{equation}
where $\sigma$ is the rms tip-tilt jitter in radians for $\sigma\ll$~1~rad. 
The relative integration time (Fig. \ref{fig:jitter}c) is given by $\tau/\tau_0=\eta_s/\eta_p^2$. The VFN configuration significantly reduces integration time with rms tip-tilt jitter that is $\ll~\lambda/D$.


\subsection{Angular size of the star}

The fundamental lower limit of $\eta_s$ is reached when the tip-tilt errors become significantly smaller than the angular size of the star. We treat the star as a ensemble of incoherent point sources in a uniform disk. For unresolved sources, $\eta_s$ may be approximated as variance of the source distribution in units of $\lambda/D$ \citep{Ruane2016dissertation}:
\begin{equation}
    \eta_s = \left(\frac{\Theta}{\sqrt{12}}\frac{D}{\lambda}\right)^2,
\end{equation}
where $\Theta$ is the angular diameter of the source. The minimum possible integration time is therefore
\begin{equation}
    \tau_\Theta = \left(\frac{\Theta}{\sqrt{12}}\frac{D}{\lambda}\right)^2\frac{\tau_0}{\eta_p^2}.
\end{equation}
More generally, $\eta_s$ may be approximated by summing the variances of the source and jitter distributions.

\subsection{Low-order aberrations in the Zernike basis}

\begin{figure*}[t!]
    \centering
    \includegraphics[width=0.75\linewidth]{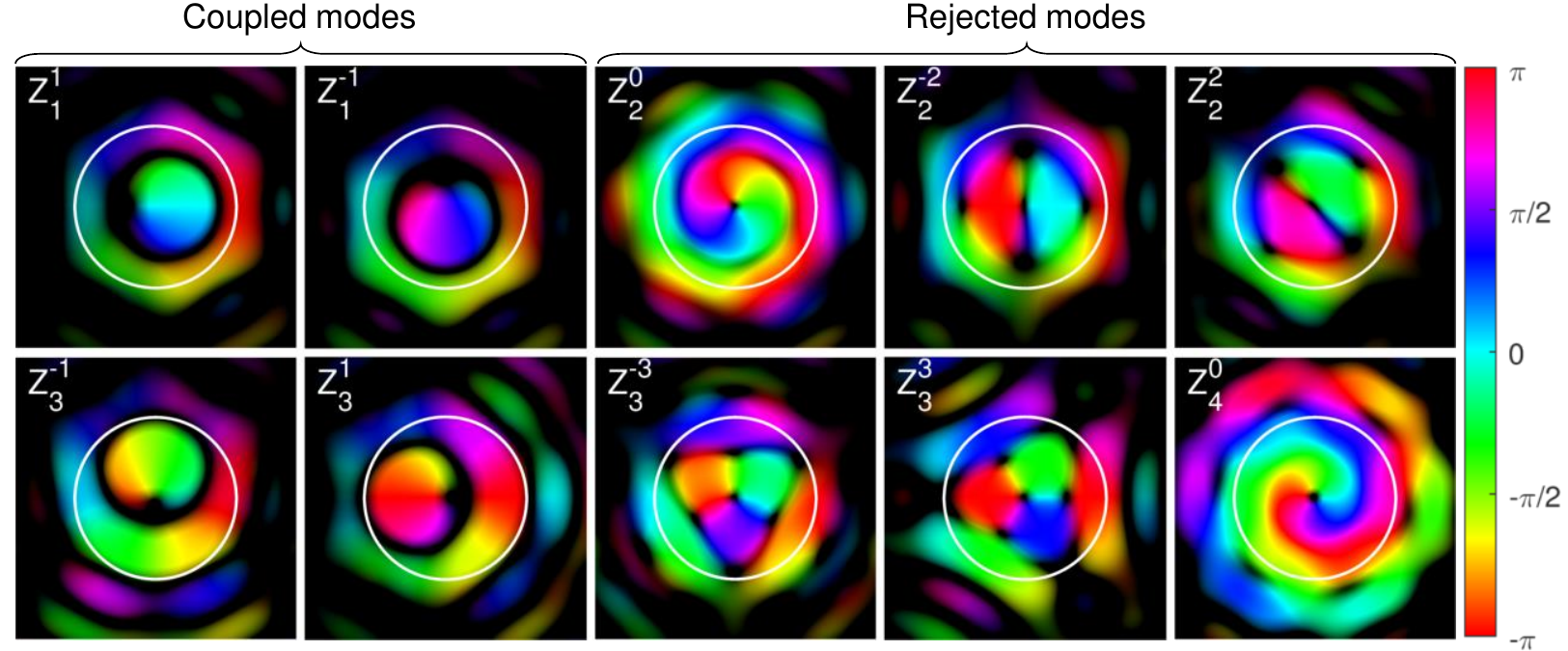}
    \caption{Amplitude (value) and phase (hue) of the stellar field in the central 4$\times$4~$\lambda F^{\#}$ region with $\lambda/10$ rms wavefront error in each of the lowest 11 Zernike modes, $Z_n^m(r,\theta)$ (excluding piston, which is shown in Fig. \ref{fig:modes}). The white circle indicates the mode field diameter $D_f=2.5~\lambda F^\#$. Modes with $|m| = 1$ leak starlight with $\eta_s\propto\omega^2$ whereas those with $|m| \ne 1$ do not couple into the SMF; i.e. $\eta_s=0$.}
    \label{fig:Zmodes}
\end{figure*}

A major benefit of VFN is that it does not rely on high-precision control of all low-order aberrations. Although an overall rms wavefront error of $\omega<1/10$ waves is desired for optimal planet light coupling, the nulling of the star is unaffected by Zernike aberrations, $Z_n^m(r,\theta)$, if $|m| \ne 1$. To provide a qualitative explanation, we write the aberrated stellar field at the pupil as a linear combination of functions of the form $R_n^m(r)e^{\pm im\theta}$, where $R_n^{m}(r)$ are the radial Zernike polynomials. Since the $\exp(\pm i\theta)$ pattern imparted on the stellar beam acts to increase the effective azimuthal order of the beam by $\pm1$, the starlight will only couple into the SMF if the the azimuthal order of the Zernike polynomial is canceled; that is, if $l+m=0$ or $l-m=0$. Thus, stellar leakage is only induced by tip-tilt and radial orders of coma, where $m = \pm 1$. 

\begin{figure*}[t!]
    \centering
    \includegraphics[width=0.325\linewidth]{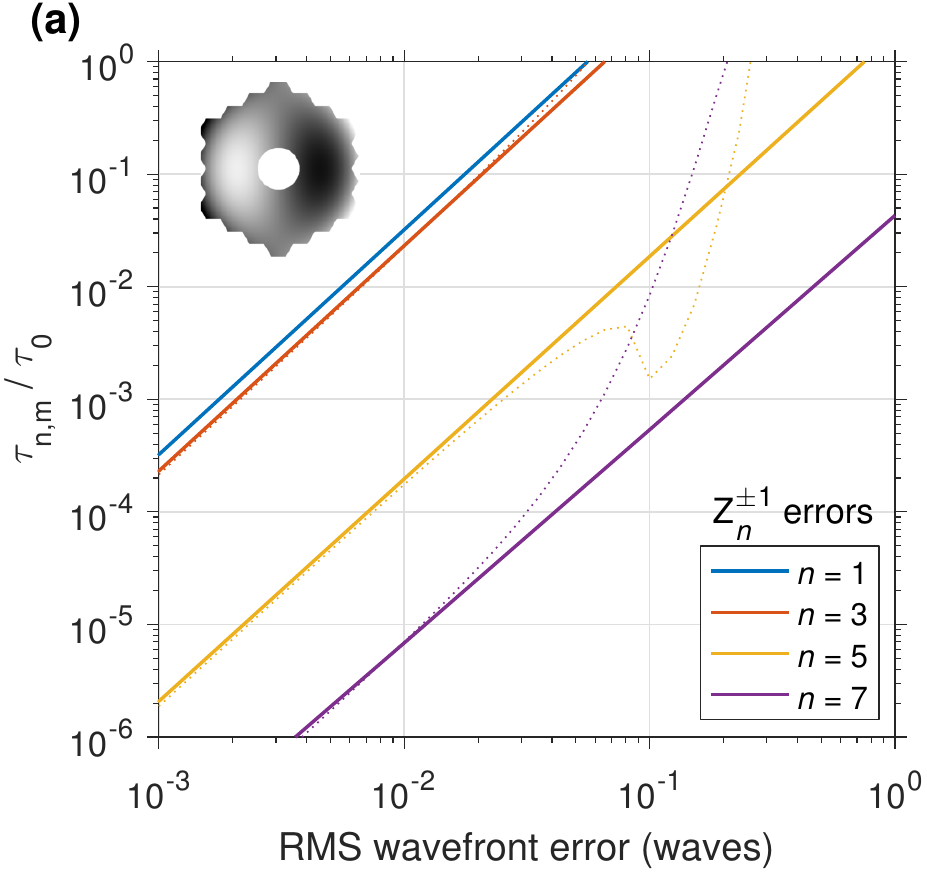}
    \includegraphics[width=0.325\linewidth]{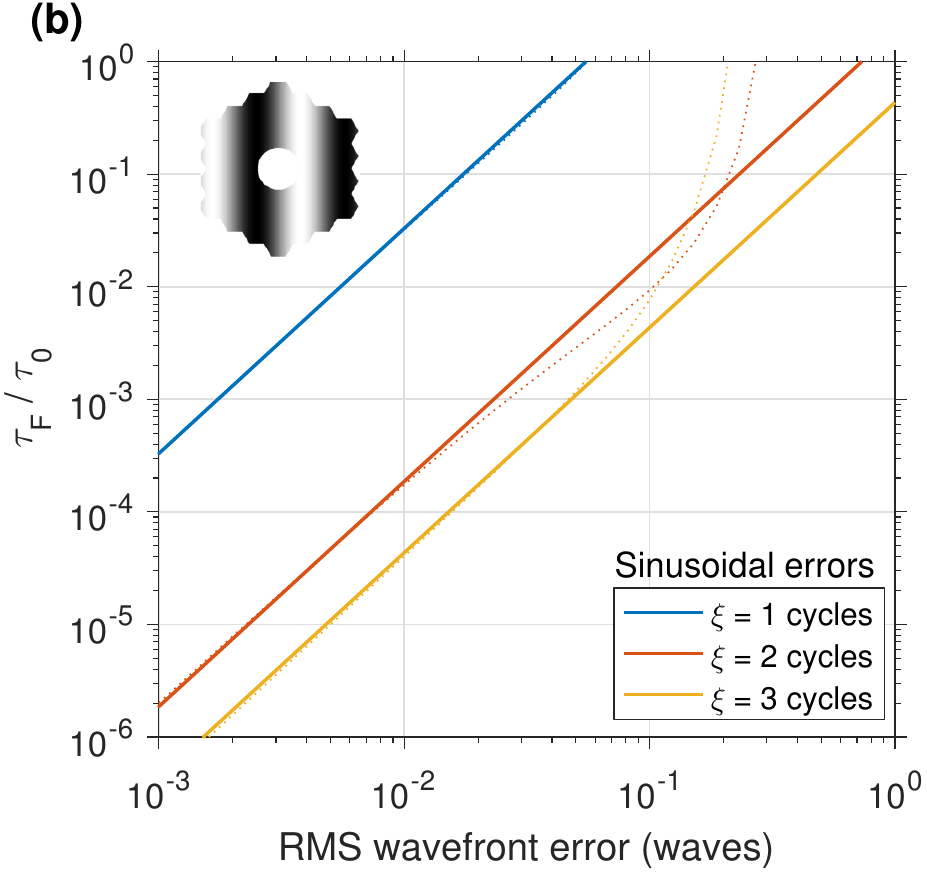}
    \includegraphics[width=0.325\linewidth]{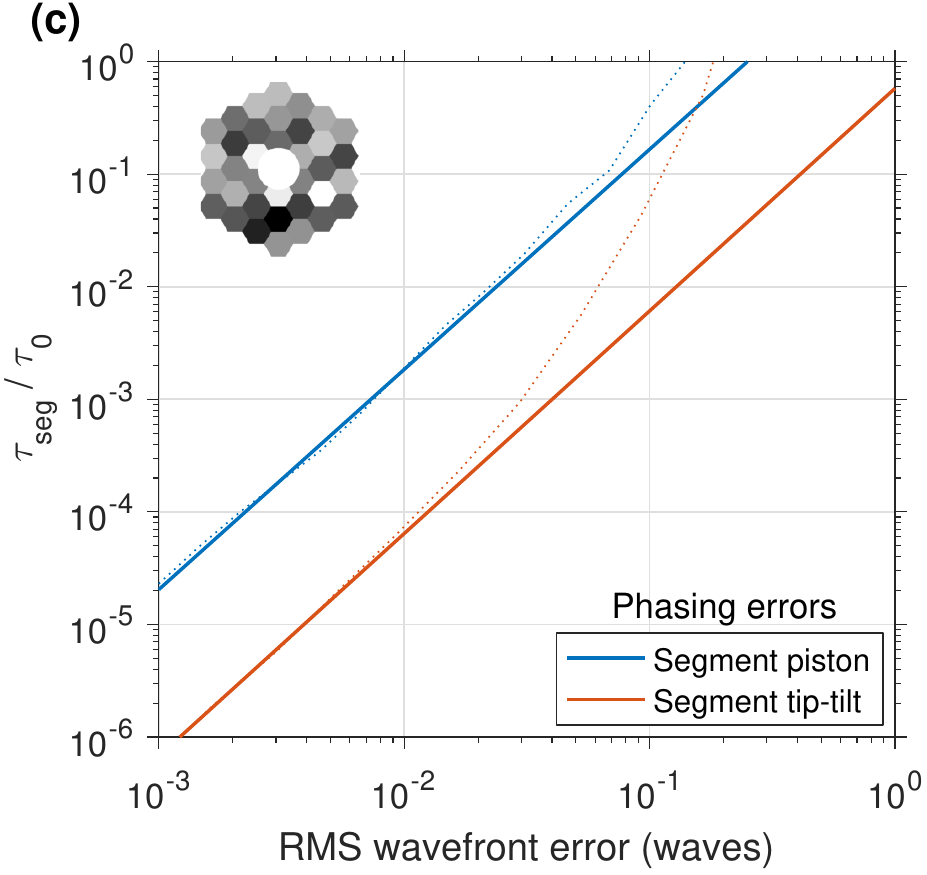}
    \caption{Sensitivity to aberrations. The relative integration time $\tau/\tau_0$ needed to overcome photon noise owing to (a)~Zernike, (b)~Fourier, and (c)~segment piston and tip-tilt aberrations. The Zernike polynomials, namely tip-tilt and ascending orders of coma, are listed by radial order $n$. In all cases, $\tau$ is well described by a second-order power law (solid lines) in the small wavefront error regime. The full numerical simulation (dotted lines) deviates from this power law when $\omega \gtrsim 1/10$ waves.}
    \label{fig:aberrations}
\end{figure*}

For small aberrations, the amount of leaked starlight is 
\begin{equation}
    \eta_s=(b_{nm}\omega)^2,
    \label{eqn:bnm}
\end{equation}
where $b_{nm}$ is the sensitivity constant. Table~\ref{tab:bs} lists numerically determined $b_{nm}$ values for the 8 lowest critical modes ($m=\pm1$). For all other modes where $|m| \ne 1$, $b_{nm}=0$ and $\eta_s=0$. Figure \ref{fig:Zmodes} illustrates this effect by plotting the phase of the field at the SMF due to an on-axis point source. The coupling is zero when the phase has a azimuthal phase ramp and/or fold symmetry. The integration time needed to overcome the noise from critical low order aberrations may be approximated by
\begin{equation}
    \tau_\mathrm{L} \approx \sum_{n,m}\tau_{nm} = \frac{\tau_0}{\eta_p^2}\sum_{n,m}(b_{nm}\omega)^2,
\end{equation}
where $\tau_{nm}$ is the integration time for each Zernike aberration. 
Figure \ref{fig:aberrations}a shows $\tau/\tau_0$ as a function of $\omega$ for the lowest order critical modes using the power-law approximation (solid lines) as well as full numerical beam propagation (dotted lines). As expected, $\tau$ has quadratic dependence on $\omega$ for $\omega\ll$1~wave, though the full simulation deviates from this behavior in some cases.

From an instrument design point of view, it is useful to budget integration time costs for dominant low order aberrations by writing the wavefront error requirements as
\begin{equation}
    \omega_\mathrm{req} = \frac{\eta_p}{b_{nm}}\sqrt{\frac{\tau_{nm}}{\tau_0}}.
\end{equation}
For example, in order to achieve $\tau/\tau_0=10^{-3}$ (i.e. with jitter $\lesssim$10$^{-2}$~$\lambda/D$~rms), the wavefront error requirements in the first three orders of coma ($Z_3^{\pm1}$, $Z_5^{\pm1}$, and $Z_7^{\pm1}$) would be $\lambda/450$, $\lambda/40$, and $\lambda/5$~rms. Since most of the power in a typical aberrated wavefront is contained in the lowest order modes, a VFN instrument should be tailored to sense and correct tip-tilt and primary coma ($Z_3^{\pm1}$) to the highest possible precision. Current AO systems provide adequate control of all other modes. 

\begin{table}
\small
\caption{Calculated low order aberration sensitivity constants (see Eqns. \ref{eqn:bnm}-\ref{eqn:bxi}). $n$ is the radial order of the Zernike polynomial (i.e. $Z_n^{\pm1}$). $\xi$ is the spatial frequency in units of cycles per pupil diameter.}
\label{tab:bs}
\begin{center}       
\begin{tabular}{c c c| c c}
\hline
\hline
\multicolumn{3}{c|}{Zernike basis} & \multicolumn{2}{c}{Fourier basis}  \\\hline
\rule[-1ex]{0pt}{3.5ex} $n$ & Noll index & $b_{n,\pm1}$ & $\xi$ & $b_\xi$\\
\hline
\rule[-1ex]{0pt}{3.5ex} 1 & 2,3 & 3.4 & 1 & 3.6\\
\rule[-1ex]{0pt}{3.5ex} 3 & 7,8 & 2.9 & 2 & 0.27 \\
\rule[-1ex]{0pt}{3.5ex} 5 & 16,17 & 0.25 & 3 & 0.13\\
\rule[-1ex]{0pt}{3.5ex} 7 & 29,30 & 0.03 & 4 & 0.13\\
\hline
\end{tabular}
\end{center}
\end{table} 

\subsection{Aberrations in the Fourier basis}

Although it is sufficient to describe the wavefront with Zernike polynomials, we also report the sensitivity to aberrations in the Fourier basis. By symmetry arguments similar to the case of low-order aberrations with $|m|\ne1$, the VFN approach is completely robust to even Fourier modes (cosines). It is, however, sensitive to odd Fourier modes (sines) with low spatial frequency. Again, we describe the amount of leaked starlight as
\begin{equation}
    \eta_s=(b_{\xi}\omega)^2,
    \label{eqn:bxi}
\end{equation}
where $b_{\xi}$ are constants that we determined numerically (see Table \ref{tab:bs}). Figure \ref{fig:aberrations}b shows the relative integration time $\tau_F/\tau_0$~=~$\eta_s/\eta_p^2$ owing to Fourier modes as a function of $\omega$ for spatial frequencies $\xi$~=~1,~2,~and~3 cycles per pupil diameter, confirming the quadratic dependence in the small error regime (solid lines). Our full simulations (dotted lines) deviate from this power law for $\omega\gtrsim 1/10$~waves~rms mainly because of a significant drop in planet coupling efficiency. 

The VFN approach is not effected by mid spatial frequency aberrations, which generate the speckle noise that tends to limit the performance of high-contrast imagers. In other words, VFN is relatively insensitive to Fourier modes with spatial frequencies $>$2-3~cycles per pupil diameter. It is therefore only necessary to control mid to high spatial frequencies to $\sim\lambda/10$~rms in order to achieve close to ideal planet coupling.

Since VFN is insensitive to high spatial frequency aberrations, the primary mirror co-phasing requirements are also somewhat relaxed for segmented telescopes. To demonstrate this, we simulated a segmented mirror with Gaussian random segment piston and tip-tilt errors (see Fig. \ref{fig:aberrations}c). Averaging over many realizations of the primary mirror, we find that VFN is most sensitive to random piston errors and is more robust to random tilts between segments. Compared to the wavefront error requirement for the primary coma term ($Z_3^{\pm1}$), the requirement for random segment pistons is relaxed by a factor of $\sim$10 for the same integration time. 

\subsection{Other noise sources}

For completeness, we include a similar analysis to determine the effect of sky and telescope thermal emission and detector noise. For thermal background noise, the integration time is given by 
\begin{equation}
    \tau_\mathrm{bg}= \frac{L_\mathrm{bg}\Omega_\mathrm{fib}}{\Phi_s}\frac{\tau_0}{\eta_p^2},
\end{equation}
where $L_\mathrm{bg}$ is the background radiance and $\Omega_\mathrm{fib}$ is the solid angle subtended by the fiber. 
For the detector dark current, $i_d$, the integration time is
\begin{equation}
    \tau_\mathrm{dc} = \frac{i_d R}{\Phi_s\lambda A q T}\frac{\tau_0}{\eta_p^2}.
\end{equation}
The read noise over long integration observations has an approximate variance of $\sigma_\mathrm{rd}^2=N_\mathrm{rd}^2 N_\mathrm{frames}\approx N_\mathrm{rd}^2 S_s/W$, where $N_\mathrm{rd}$ is the read noise, $N_\mathrm{frames}$ is the number of frames, and $W$ is the full well depth. Thus, the integration time to overcome read noise is
\begin{equation}
    \tau_\mathrm{rd} = \frac{N_\mathrm{rd}^2}{W}\frac{\eta_s}{\eta_p^2}\tau_0,
\end{equation}
which depends on $\eta_s$ and, by extension, on the estimated optical aberrations present. 

\begin{table*}[t]
\small
\caption{Optical performance of a VFN instrument for the case of Ross~128~b as a function of wavelength and telescope size. The angular separation between the planet and star is 15~mas and the star has an angular diameter of 0.5~mas. $\eta_p$ is calculated at the central wavelength.}
\label{tab:finitestars}
\begin{center}       
\begin{tabular}{c|c c c|c c c|c c c}
\hline\hline
Tel. Diam:  & \multicolumn{3}{c|}{10~m} & \multicolumn{3}{c|}{15~m} & \multicolumn{3}{c}{30~m}\\\hline
\rule[-1ex]{0pt}{3.5ex} Band& $\lambda/D$ & min.~$\eta_s$ & $\eta_p$ & $\lambda/D$ & min.~$\eta_s$ & $\eta_p$ & $\lambda/D$ & min.~$\eta_s$ & $\eta_p$ \\
\hline
\rule[-1ex]{0pt}{3.5ex} $r$~(0.77$\mu$m) & 16~mas & 1$\times$10$^{-4}$ & 0.20  & 11~mas & 2$\times$10$^{-4}$ & 0.10 & 5~mas & 9$\times$10$^{-4}$ & $<$0.01 \\
\rule[-1ex]{0pt}{3.5ex} $J$~(1.2$\mu$m) & 26~mas & 4$\times$10$^{-5}$ & 0.15  & 17~mas & 8$\times$10$^{-5}$ & 0.20 & 9~mas & 3$\times$10$^{-4}$ & 0.02 \\
\rule[-1ex]{0pt}{3.5ex} $H$~(1.6$\mu$m) & 34~mas & 2$\times$10$^{-5}$ & 0.10  & 22~mas & 5$\times$10$^{-5}$ & 0.17 & 11~mas & 2$\times$10$^{-4}$ & 0.12  \\
\rule[-1ex]{0pt}{3.5ex} $K$~(2.2$\mu$m) & 45~mas & 1$\times$10$^{-5}$ & 0.06  & 30~mas & 3$\times$10$^{-5}$ & 0.12 & 15~mas & 1$\times$10$^{-4}$ & 0.20 \\
\rule[-1ex]{0pt}{3.5ex} $L^\prime$~(3.8$\mu$m) & 78~mas & 4$\times$10$^{-6}$ & 0.02  & 52~mas & 9$\times$10$^{-6}$ & 0.05 & 26~mas & 4$\times$10$^{-5}$ & 0.15 \\
\hline

\end{tabular}
\end{center}
\end{table*} 

\subsection{Estimating total integration time}

Since the variance of each noise term has been approximated to be linear with integration time, the total integration time may be estimated by summing the integration time to overcome each independent noise source:
\begin{equation}
    \tau = \tau_\mathrm{tt} + \tau_\Theta + \tau_L + \tau_\mathrm{bg} + \tau_\mathrm{dc} + \tau_\mathrm{rd},
\end{equation}
which is equivalent to summing the individual noise variances. With this expression, it is straightforward to predict the feasibility of characterizing planets with a theoretical VFN instrument and to develop engineering requirements.

\section{Requirements for characterizing Ross 128 \MakeLowercase{b}}

In this section, we investigate the feasibility of detecting and characterizing the recently discovered planet Ross~128~b with a VFN instrument on Keck, TMT, and LUVOIR. Recently detected by \citet{Bonfils2017} using the radial velocity (RV) measurements, Ross~128~b has a minimum mass of 1.27~$M_\Earth$ and an orbital semi major axis of 0.05~au, which is within the habitable zone of the cool ($T_\mathrm{eff}=3200~K$) M4V dwarf located at 3.38~pc. Ross~128 is representative of potentially habitable planet host stars that may be targeted with future VFN instruments on large-aperture ground- and space-based telescopes. We first determine the required SNR per spectral channel and then calculate the integration time needed to detect the planet in the presence of stellar photon noise as a function of the telescope size, wavelength, and wavefront aberrations. 

\subsection{Required SNR per spectral channel}

Detecting molecules in the atmosphere of Ross~128~b with a VFN will require cross-correlating the measured high-resolution spectrum with a template spectrum in order to make efficient use of the relatively few planet photons and the spectral information they carry. A peak in the cross-correlation function (CCF) indicates the presence of a planet with a spectrum similar to the template. In practice, the planet is characterized by modeling the spectra of planets with a variety of possible compositions and finding the template that maximizes the peak in the CCF \citep{Wang2017_HDCI,Hoeijmakers2018}. 

Using the methods developed in \citet{Wang2017_HDCI}, we determine the relationship between the SNR of the peak in the CCF and the SNR per spectral channel for an R=100,000 spectrograph. The spectrum and matching template for an Earth-like planet covered by low clouds orbiting an M dwarf is simulated using an atmospheric chemistry and radiative transfer model \citep{Hu2012a,Hu2012b,Hu2013,Hu2014}. The resulting geometric albedo is $\sim$0.1. Assuming a planet radius of 1.5$R_\Earth$, the planet-to-star flux ratio in reflected light is $\sim$5$\times10^{-8}$. The assumed stellar spectrum is from the PHOENIX BT-Settl model grid \citep{Baraffe2015}.

\begin{figure}[t!]
    \centering
    \includegraphics[width=0.85\linewidth]{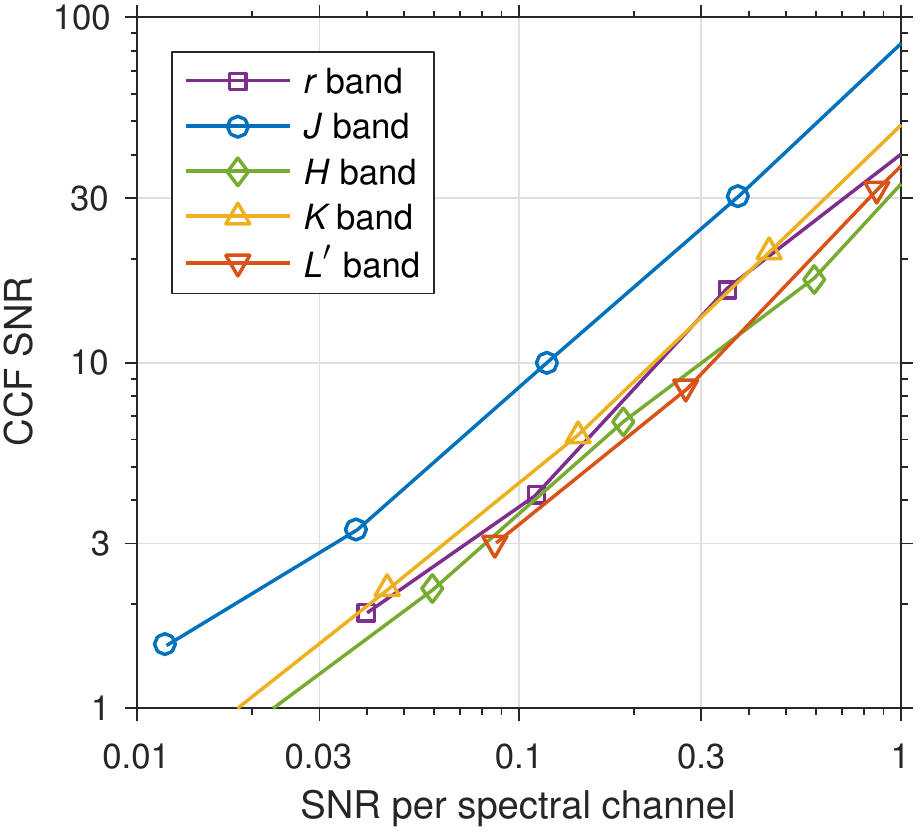}
    \caption{Signal-to-noise ratio (SNR) of the peak in the cross-correlation function (CCF) as a function of the SNR per spectral channel from simulated observations in $r$, $J$, $H$, $K$, and $L^\prime$ bands, using low-cloud Earth spectrum and the simulations methods developed in \citet{Wang2017_HDCI}. The required SNR per spectral channel for a CCF SNR of 5 in $J$ band is 0.06, whereas 0.11-0.16 is required in the remaining bands. The actual values are listed in Table \ref{tab:tau_Ross128b_finiteStar}.}
    \label{fig:SNRgain}
\end{figure}

We calculate the distribution of peak values of the resulting CCFs as a function of the SNR per spectral channel for 100 observations in a set of wavelength ranges that correspond to $r$, $J$, $H$, $K$, and $L^\prime$ bands (see Table~\ref{tab:finitestars}). In practice, the number of pixels in the spectrograph or atmospheric dispersion will likely limit the maximum spectral bandwidth. Assuming that stellar photon noise is the dominant noise source and that the planet spectrum is known, Figure \ref{fig:SNRgain} shows the resulting CCF SNR, defined as the ratio between the median of the peak values and the standard deviation of the distribution of peak values. We find that the ratio between the CCF SNR and the SNR per spectral channel is the highest for $J$ band. The required SNR per spectral channel for a CCF SNR of 5 in $J$ band is 0.06, whereas 0.11-0.16 is required in the remaining bands (see Table \ref{tab:tau_Ross128b_finiteStar}). These values are higher when the template spectrum differs from the true planet spectrum, as discussed in \citet{Wang2017_HDCI}.

\begin{table}[t]
\small
\caption{The signal to noise ratio (SNR) per spectral channel and minimum integration time needed to detect Ross~128~b in $r$, $J$, $H$, $K$, and $L^\prime$ bands at a CCF SNR of 5 ignoring detector noise, background noise, and wavefront aberrations. This fundamental limit is set by stellar photon noise due to the diameter of the star (0.5~mas). Whereas $J$ band wavelengths are optimal for observing Ross~128~b with Keck and LUVOIR, the integration time is smaller at longer wavelengths ($K$ and $L^\prime$ bands) on TMT.}
\label{tab:tau_Ross128b_finiteStar}
\begin{center}       
\begin{tabular}{c c|c c c}
\hline\hline
\multicolumn{2}{c|}{Goal} & \multicolumn{3}{c}{Minimum $\tau$ (hr)} \\\hline
\rule[-1ex]{0pt}{3.5ex} Band & SNR & Keck & LUVOIR & TMT \\
\hline
\rule[-1ex]{0pt}{3.5ex} $r$ & 0.13 & 3260 & 14718 & - \\
\rule[-1ex]{0pt}{3.5ex} $J$ & 0.06 & 53 & 31 & 2125 \\
\rule[-1ex]{0pt}{3.5ex} $H$ & 0.14 & 356 & 136 & 261 \\
\rule[-1ex]{0pt}{3.5ex} $K$ & 0.11 & 429 & 126 & 45 \\
\rule[-1ex]{0pt}{3.5ex} $L^\prime$ & 0.16 & 1970 & 469 & 50 \\
\hline
\end{tabular}
\end{center}
\end{table} 


\begin{table}[t!]
\small
\caption{Parameters used in integration time calculations. We adopt the detector noise characteristics for a science grade Teledyne H2RG HgCdTe focal plane array cooled to 77~K \citep{Blank2012} and transmission estimates based on the KPIC instrument. The majority of Ross 128 b parameters are from \citet{Bonfils2017}. The planet radius and albedo are currently not well constrained.}
\label{tab:simparams}
\begin{center}       
\begin{tabular}{l c}
\hline\hline
\rule[-1ex]{0pt}{3.5ex} \textbf{Instrument parameters} & \\
\hline
\rule[-1ex]{0pt}{3.5ex} Telescope transmission, $T$ & 0.3\\
\rule[-1ex]{0pt}{3.5ex} Telescope diameter, $D$ [m] & 10, 15, or 30 \\
\rule[-1ex]{0pt}{3.5ex} Collecting area, $A$ [m$^2$] & 76, 157, or 655 \\
\rule[-1ex]{0pt}{3.5ex} Mode field diameter, $D_f$ [$\lambda F^\#$] & 2.5 \\
\rule[-1ex]{0pt}{3.5ex} Spectral resolution, $R$ & 100,000\\
\rule[-1ex]{0pt}{3.5ex} Spectral bandwidth, $\Delta\lambda/\lambda$ & $\sim$0.2\\
\rule[-1ex]{0pt}{3.5ex} Detective quantum efficiency & 0.85 \\
\rule[-1ex]{0pt}{3.5ex} Dark current, $i_\mathrm{d}$ [e$^{-}$/sec] & 0.002 \\
\rule[-1ex]{0pt}{3.5ex} Read noise, $N_\mathrm{rd}$ [e$^{-}$] & 3.2 \\
\rule[-1ex]{0pt}{3.5ex} Full well depth [e$^{-}$] & 109,000 \\
\hline
\rule[-1ex]{0pt}{3.5ex} \textbf{Ross 128 (star) parameters} & \\\hline
\rule[-1ex]{0pt}{3.5ex} Distance [pc] &  3.381\\
\rule[-1ex]{0pt}{3.5ex} Radius, $r_s$ [$R_\odot$] & 0.1967 \\
\rule[-1ex]{0pt}{3.5ex} Temperature, $T_s$ [K] & 3192 \\
\rule[-1ex]{0pt}{3.5ex} Angular size, $\Theta$ [mas] & 0.5 \\\hline
\rule[-1ex]{0pt}{3.5ex} \textbf{Ross 128 b (planet) parameters} & \\\hline
\rule[-1ex]{0pt}{3.5ex} Semi-major axis, $a_p$ [au] & 0.0496\\
\rule[-1ex]{0pt}{3.5ex} Radius, $r_p$ [$R_\oplus$] & 1.5 \\
\rule[-1ex]{0pt}{3.5ex} Temperature, $T_p$ [K] & 280 \\
\rule[-1ex]{0pt}{3.5ex} Angular separation [mas] & 15\\
\rule[-1ex]{0pt}{3.5ex} Albedo, $\alpha$ & 0.1\\
\rule[-1ex]{0pt}{3.5ex} Phase function, $\phi$ (max. elongation) & 0.3\\
\hline
\end{tabular}
\end{center}
\end{table}

\subsection{Keck telescope}

The 10~m Keck telescope is an ideal platform to demonstrate VFN, paving the way for the characterization of Earth-sized planets with future telescopes, such as LUVOIR and TMT. Furthermore, VFN may be implemented as part of the fiber injection unit (FIU) of the Keck Planet Imager and Characterizer (KPIC) with minimal modification to the current instrument design \citep{Mawet2016_KPIC,Mawet2017_KPIC}. 

Ross~128~b has an angular separation of $\sim$15~mas at maximum elongation, which determines $\eta_p$ as a function of wavelength for the given telescope size. The minimum possible $\eta_s$ is set by the angular diameter of the star: $\sim$0.5~mas. With the VFN performance parameters listed in Table~\ref{tab:finitestars}, we computed the minimum integration time for detecting Ross~128~b at a CCF SNR of 5 with the template-matching technique. In cases that require more than $\sim$6~hr of integration time, the full observing campaign is made up of the summation of a number of six-hour observations taken on multiple nights. The discrete observations are planned to coincide with the planet's maximum elongation. About that point, assuming Ross~128~b is in an edge-on circular orbit, the angular separation only changes by 0.3\% in six hours and therefore has a negligible effect on the integration time estimate.

\begin{figure}[t!]
    \centering
    \includegraphics[width=0.85\linewidth]{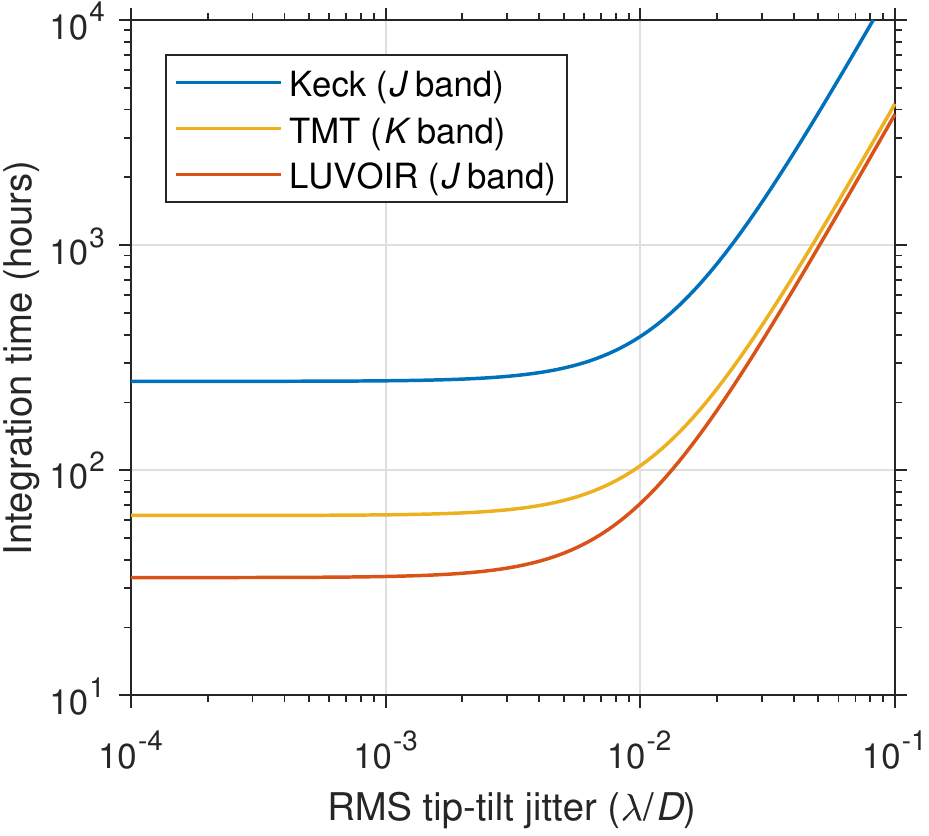}
    \caption{Integration time required for achieving a CCF SNR of 5 on Ross~128~b as a function of tip-tilt wavefront errors. We assume 5~nm~rms of coma for the ground-based telescopes (Keck and TMT) and 1~nm~rms for space telescopes (LUVOIR). Other relevant assumptions are given in Table \ref{tab:simparams}.}
    \label{fig:tau_vs_tt_Ross128b}
\end{figure}

The combined effect of the VFN, the late spectral type of the star, and the SNR gain achieved with template-matching make $J$ band the optimal wavelength range for observing Ross~128~b with Keck (see Table \ref{tab:tau_Ross128b_finiteStar}). Spectroscopy in $J$ band also provides a means to measure the abundance of H$_2$O, O$_2$, and CO$_2$ which are key to assessing habitability \citep{Meadows2018}. The integration times in Table \ref{tab:tau_Ross128b_finiteStar} assume the planet spectrum is known and do not include the effect of wavefront error, which is the dominant contributor to $\eta_s$. 
Figure~\ref{fig:tau_vs_tt_Ross128b} shows an estimate of the integration time for achieving a CCF SNR of 5 as function of rms tip-tilt jitter, assuming coma aberrations are controlled to 5~nm~rms with Keck AO or an upgraded version thereof. The remaining assumptions are listed in Table \ref{tab:simparams}. Controlling tip-tilt to $10^{-2}$~$\lambda/D$~rms and coma to 5~nm~rms yields an integration time of $\sim$400~hr.

Current state-of-the-art AO systems  \citep[e.g. SCExAO;][]{Jovanovic2015} are capable of controlling tip-tilt errors to $\sim10^{-2}$~$\lambda/D$~rms and low-order Zernike modes to $\sim$50~nm~rms \citep{Singh2015}. Although the wavefront error requirements shown in Fig.~\ref{fig:tau_vs_tt_Ross128b} are 10$\times$ smaller than currently achieved on SCExAO, improved correction is expected with closed-loop predictive control.  \citet{Males2018} recently showed that starlight suppression may be improved by a factor of 1400$\times$ for high-contrast imaging with a coronagraph and a bright guide star, which corresponds to a reduction in low-order wavefront error by a factor of approximately $\sqrt{1400}=37\times$. Significant reductions in low-order wavefront error are especially feasible for a VFN instrument since only a single mode (coma) needs to be corrected to higher precision than current AO systems provide.

\subsection{Thirty Meter Telescope (TMT)}

The angular separation of Ross~128~b is approximately the VFN outer working angle with TMT at $H$ band where the throughput begins to drop rapidly (see Table~\ref{tab:finitestars}). On the other hand, thermal background noise (not included in Table \ref{tab:tau_Ross128b_finiteStar}) dominates the photon noise budget at $L^\prime$ and longer wavelengths. Thus, $K$ band is the optimal wavelength range for observing Ross~128~b with VFN on TMT. 

Provided the same AO performance is achieved with TMT as assumed above for Keck (i.e. $10^{-2}$~$\lambda/D$~rms tip-tilt jitter and 5~nm~rms of coma), Fig. \ref{fig:tau_vs_tt_Ross128b} shows that the integration time needed to detect Ross~128~b with a CCF SNR of 5 in $K$ band is 100~hr, approximately four times less than with Keck in $J$ band. 

$K$-band observations with TMT will be especially important for detecting CH$_4$. Additionally, $K$-band observations will constrain the abundance of H$_2$O and CO$_2$, but are not sensitive to O$_2$. Therefore, conventional coronagraph observations at $J$ band will complement VFN observations at $K$ band. In fact, we confirmed that conventional coronagraphs provide shorter integration times than VFN in $J$ band and at smaller wavelengths: for instance, using a vortex coronagraph and FIU \citep{Mawet2017_HDCII}, due to the improvement in throughput that the coronagraph provides for angular separations $>\lambda/D$.

\subsection{LUVOIR space telescope}

The 15~m LUVOIR space telescope may be the most efficient platform for characterizing Earth-sized planets in the habitable zone of M dwarfs with a VFN instrument. Since imaging Earth-like planets around sun-like stars is a premier science goal of LUVOIR, the telescope already has strict stability requirements and extremely precise wavefront control capability ($<$10~pm~rms). The integration time for detecting Ross~128~b would likely reach the fundamental limit set by the finite size of the star. 

Similar to the Keck telescope, the optimal wavelength range for observing Ross~128~b with LUVOIR is $J$ band. With the assumptions in Table \ref{tab:simparams}, a CCF SNR of 5 is achieved in only 30~hr of integration. As mentioned above in the context of Keck observations, $J$-band VFN observations also provide the opportunity to measure the abundance of potential biosignatures H$_2$O, O$_2$, and CO$_2$. 

Compared to imaging Earth-like planets in the habitable zone of sun-like stars, where a planet-to-star flux ratio of $\epsilon=10^{-10}$ is expected, VFN observations of M dwarfs with LUVOIR will not require the time-consuming wavefront control calibrations to dig a dark hole in the speckle pattern at the coronagraph image plane. VFN would thereby be an efficient method for surveying nearby M dwarfs to detect new planets and following up known planets detected through direct imaging at shorter wavelengths, RV measurements, or Gaia astrometry. 

\subsection{Uncertainty in integration time calculations}

A major source of uncertainty in the Ross~128~b integration time calculations is that the planet radius, $r_p$, is poorly constrained by its RV detection. Since the integration time scales inversely as the square of the flux ratio ($\tau\propto\epsilon^{-2}$) and the flux ratio scales as the square of the planet radius ($\epsilon\propto r_p^{2}$), the dependence of integration time on planet radius is a fourth-order power-law relationship: $\tau\propto r_p^{-4}$. For instance, a planet that is twice the size would take 1/16th of the amount of time to detect. The measured minimum mass of 1.27~$M_\Earth$ implies that Ross~128~b is slightly larger than Earth. We adopted $r_p=1.5 R_\oplus$ in the calculations above. However, assuming the orbit is edge-on and assuming the planet has the same density as Earth, the radius of the planet is $r_p=(1.26)^{1/3} R_\oplus=1.1 R_\oplus$ and the integration time would be 3.4$\times$ greater. 

Likewise, the planet albedo, $\alpha$, is also unknown. We assumed a value of $\alpha\approx0.1$ based on a low-cloud atmospheric chemistry and a radiative transfer model \citep{Hu2012a,Hu2012b,Hu2013,Hu2014}. However, we find that the albedo may be five times greater for a high-cloud model. Since the integration time scales as $\tau\propto \alpha^{-2}$, the integration time could be up 25$\times$ shorter with high clouds. On the other hand, the spectral lines differ considerably between low-cloud and high-cloud models, which alters relationship between the CCF SNR and SNR per spectral channel, introducing additional uncertainty.  

The assumed orbital inclination affects both the assumed mass and the orbital phase function, $\phi$. However, the true mass could be much greater, in which case, the planet may be larger in size and detected in a much shorter integration time. In addition, the integration time is inversely proportional to the phase function, $\tau\propto \phi^{-2}$, and therefore the integration time can vary by a factor of $\sim$10 depending on the planet's orbital inclination and the position along its orbit. 

Another source of uncertainty is the performance characteristics of future telescopes and instruments. In the calculations above, we assume a transmission of $T=0.3$ (including the telescope and spectrograph) and detector quantum efficiency of $q=0.85$. The former is based on end-to-end transmission estimates of the Keck telescope and the KPIC instrument. The latter is representative of a Teledyne H2RG HgCdTe focal plane array \citep{Blank2012}. However, these values can potentially be higher in a dedicated VFN instrument with fewer reflections, higher throughput spectrographs, and improved detectors. The integration time is inversely proportional to each of these quantities.

\section{Discussion}

\subsection{Detecting new Earth-like planets orbiting M stars}

The approach outlined for the Ross~128~b case above may potentially be used to detect new Earth-like planets in the habitable zone of other nearby M stars. We estimated the number of stars around which a CCF SNR of 5 could be achieved for an Earth-sized planet at the Earth isolation distance with the same spectrum, template, and noise properties used above. Using a VFN instrument mode in $J$ band, 10, 15, and 30~m telescopes can respectively search 3, 7, 45 habitable zones for such planets in less than 50 hr. Whereas the targets for a 10-meter telescope are within 3~pc, the habitable zones accessed by a 30-meter are around stars as far as 8~pc.

\subsection{New scientific opportunities}

In addition to enabling the characterization of Earth-sized planets in the habitable zone of nearby M dwarfs, a VFN instrument would enable new scientific opportunities for current and future space telescopes. 

Upgrading the FIU of KPIC \citep{Mawet2016_KPIC,Mawet2017_KPIC} would simply require a new focusing lens. 
This new mode would allow for the detection of all planet types. Blind surveys of nearby stars may yield new detections of giant planets on close-in orbits and the chemical makeup of their atmospheres. However, a more effective approach is to perform targeted characterization of known planets detected via direct imaging at shorter wavelengths, RV \citep[with e.g. SPIRou;][]{Artigau2014}, and astrometry \citep[with e.g. Gaia;][]{Perryman2014}. The high-resolution spectrum also provides constraints on the orbital radial velocity \citep{Snellen2014} and, with sufficient SNR, the rotation rate \citep{Bryan2017}. Larger aperture telescopes, such as TMT, will extend the above possibilities to planets with smaller $\epsilon$ values and angular separations. 



\subsection{Potential design improvements}

Although the $F^\#$ of the system has been chosen to maximize the coupling of planet light into the SMF, it is possible that varying the $F^\#$ may lead to lower integration times under a given set of wavefront error assumptions \citep{Mennesson2002}. In addition, beam shaping techniques, such as phase induced amplitude apodization, may be a pathway to improve throughput \citep{Jovanovic2017}. 

Efficient VFN observations require very tight constraints on tip-tilt and coma wavefront errors. Fortunately, future VFN instruments can take advantage of recent developments in low-order wavefront sensing and control, including reflective Lyot stop \citep{Singh2015} and holographic wavefront sensors designed to sense the $\exp(\pm i\theta)$ modes upstream of the vortex phase mask \citep{Wilby2017}. All of the above will benefit from closed-loop predictive control under development on state-of-the-art AO systems \citep{Males2018}.






With low-noise, high-speed infrared detectors \citep[e.g.][]{Greffe2016}, it is possible to obtain spectra at high enough frame rates ($>$kHz) to prevent averaging over the full distribution of tip-tilt and coma errors. Excluding frames where the errors are greater than a given threshold may reduce the effective $\eta_s$ and thereby the integration time needed for detection. A similar approach was presented in \citet{Hanot2011}.

\subsection{Limitations}

VFN is fundamentally limited in terms of outer working angle. Making $\lambda F^\#$ smaller with respect to the fiber mode incurs significant throughput losses. It is more efficient to characterize planets outside of $\lambda/D$ with point spectroscopy downstream of a conventional high-contrast imager. 

Another major limitation for VFN is that the integration time needed to characterize planets at $\epsilon<10^{-8}$ becomes longer than a feasible observing program. For example, the signal from Earth-like planets in the habitable zone of sun-like stars with $\epsilon\approx10^{-10}$ will be overwhelmed by stellar photon noise owing to the finite size of the star ($\sim$1~mas) and would require $>$1000$\times$ longer integration times than Ross~128~b. That science case calls for a larger inner working angle coronagraph, such as the vortex coronagraphs proposed for the Habitable Exoplanet Imaging Mission \citep[HabEx;][]{Mennesson2016,Ruane2017_JATIS}.


\section{Conclusion}

The vortex fiber-nulling (VFN) method uses the combination of a vortex phase mask and single-mode fiber to reject starlight, while feeding light from planets at angular separations of $\lesssim\lambda/D$ to a spectrograph. Combined with the template-matching technique at high spectral resolution, VFN provides a pathway to characterize Earth-sized planets in the habitable zone of nearby M dwarfs, such as Ross~128~b, in $\sim$400, $\sim$100, and $\sim$30~hr with Keck, TMT, and LUVOIR, respectively. The integration time is strongly dependent on the planet properties as well as tip-tilt and coma aberrations but is otherwise insensitive to wavefront errors. VFN requires wavefront residuals of $\sim10^{-2}$~$\lambda/D$~rms of tip-tilt and on the order of a few nm~rms of coma. The finite size of the star dominates the stellar photon noise budget in the case for ultra-stable space telescopes designed for exoplanet imaging, such as LUVOIR. VFN enables efficient, targeted characterization of known planets orbiting nearby stars. A laboratory demonstration of the VFN technique is underway at Caltech's Exoplanet Technology (ET) Laboratory \citep{Delorme2018}. 

\acknowledgments

G. Ruane is supported by an NSF Astronomy and Astrophysics Postdoctoral Fellowship under award AST-1602444.




\bibliography{RuaneLibrary}   
\bibliographystyle{aasjournal}   

\end{document}